\documentclass[12pt]{spieman}  % 12pt font required by SPIE;
\usepackage{amsmath,amsfonts,amssymb}
\usepackage{graphicx}
\usepackage{setspace}
\usepackage{tocloft}
\usepackage{lineno}

\title{Efficient wavefront sensing for space-based adaptive optics}

\author[a, c*]{He Sun}
\author[a]{N. Jeremy Kasdin}
\author[b]{Robert Vanderbei}
\affil[a]{Department of Mechanical and Aerospace Engineering, Princeton University, NJ, the United States, 08540}
\affil[b]{Department of Operation Research and Financial Engineering, Princeton University, NJ, the United States, 08540}
\affil[c]{Department of Computing and Mathematical Sciences, California Institute of Technology, CA, the United States, 91125}

\cftpagenumbersoff{figure}
\cftpagenumbersoff{table} 
\begin{document} 
\maketitle

\begin{abstract}
Future large space telescopes will be equipped with adaptive optics (AO) to overcome wavefront aberrations and achieve high contrast for imaging faint astronomical objects, such as earth-like exoplanets and debris disks. In contrast to AO that is widely used in ground telescopes, space-based AO systems will use focal plane wavefront sensing to measure the wavefront aberrations. Focal plane wavefront sensing is a class of techniques that reconstruct the light field based on multiple focal plane images distorted by deformable mirror (DM) probing perturbations. In this paper, we report an efficient focal plane wavefront sensing approach for space-based AO which optimizes the DM probing perturbation and thus also the integration time for each image. Simulation of the AO system equipped with a vortex coronagraph has demonstrated that our new approach enables efficient information acquisition and significantly reduces the time needed for achieving high contrast in space.
\end{abstract}

% Include a list of up to six keywords after the abstract
\keywords{adaptive optics, high-contrast imaging, coronagraph, exoplanet, optimal experiment design}

% Include email contact information for corresponding author
{\noindent \footnotesize\textbf{*}He Sun,  \linkable{hesun@caltech.edu} }

\begin{spacing}{1}   % use double spacing for rest of manuscript
%\linenumbers

\section{Introduction} \label{sec:intro}
\begin{figure}[h!]
\centering\includegraphics[width=14cm]{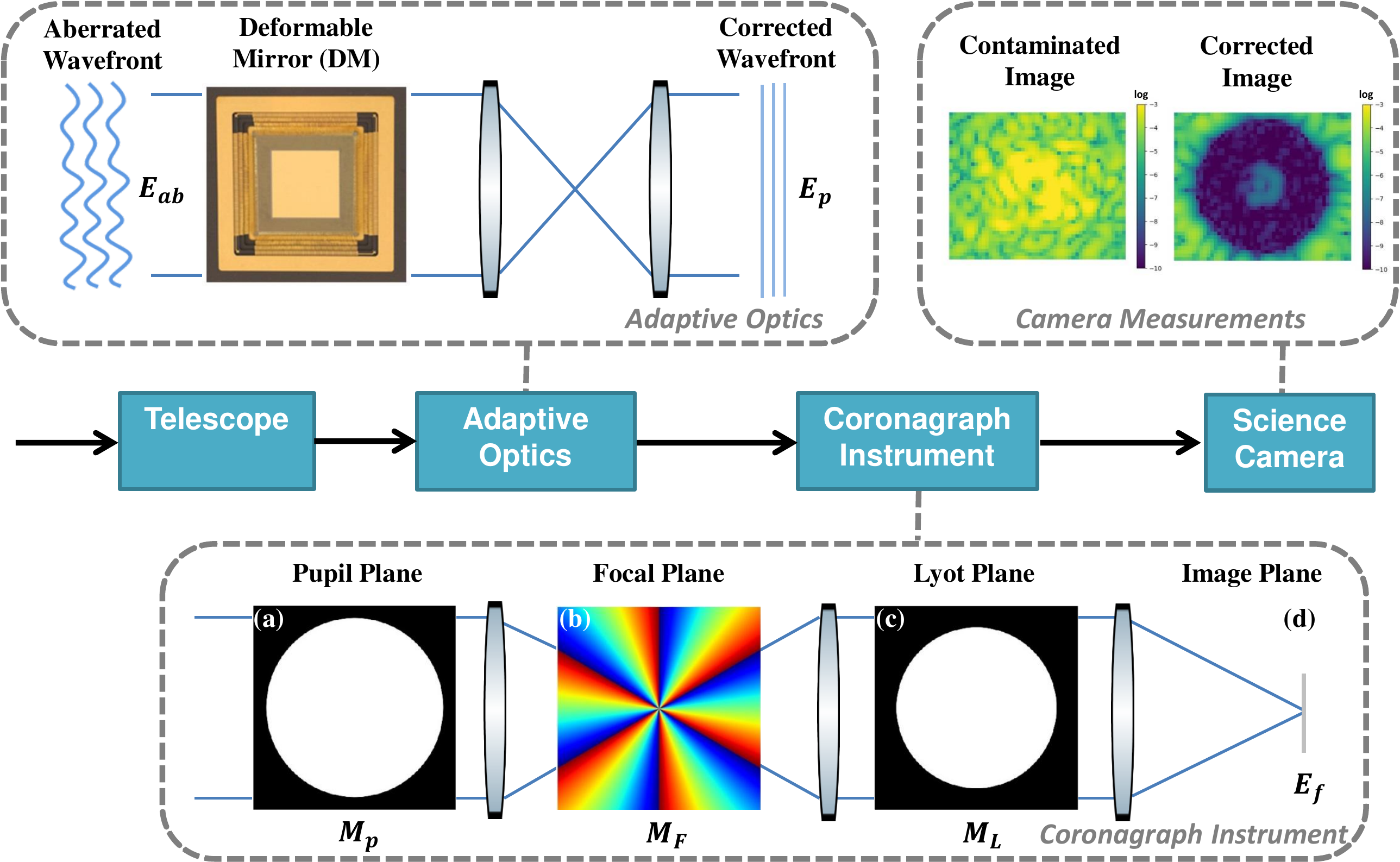}
\caption{A space telescope system equipped with AO and a high-contrast vortex coronagraph. The AO corrects \textcolor{black}{the complex wavefront aberrations \textcolor{black}{(only phase aberrations are shown here)} using deformable mirrors (only one mirror is shown here for simplicity, however, typically more than two deformable mirrors are used in a real AO system)} and the coronagraph suppresses the starlight using a series of masks, including (a) \textcolor{black}{pupil plane mask} (a binary mask, fully transmissive or fully opaque), (b) a focal plane vortex mask\cite{mawet2009vector} (a pure phase mask, its phase shown in the figure), and (c) a Lyot stop (a binary mask). After several AO control steps, a high-contrast annular observation region appears in the image.}
\label{fig:AO}
\end{figure}

One of the major goals for the next-generation large space telescopes\cite{spergel2015wide, mennesson2016habitable, bolcar2017large} is to directly image faint earth-like planets. This requires that the  telescope be equipped with a coronagraph\cite{kasdin2003extrasolar, guyon2003phase, mawet2009vector, zimmerman2016shaped, trauger2016hybrid} for suppressing starlight and adaptive optics (AO) for correcting wavefront aberrations (see Fig.~\ref{fig:AO}). In ground-based telescopes\cite{guyon2010subaru, macintosh2014first}, adaptive optics typically works by first measuring the wavefront aberrations using a wavefront sensor\cite{verinaud2005adaptive} and then compensating for the aberrations using devices such as deformable mirrors (DMs)\cite{bifano2011adaptive}. However, this conventional approach is not suitable for space missions where a higher contrast (below $10^{-9}$ rather than the $10^{-6}$ typical of ground telescopes) is required because a separate wavefront sensor introduces non-common-path errors. Instead, focal plane wavefront sensing\cite{sun2019modern} must be used in space-based AO to retrieve the aberrated light field.  This is done via small probing commands to the DMs, causing the light field to vary slightly, allowing it to be estimated by observing the corresponding focal plane intensity changes and solving a phase-retrieval optimization problem.

Currently, the benchmark method for focal plane wavefront sensing is pair-wise DM probing followed by a batch process estimation\cite{borde2006high, give2011pair}. This approach constructs a linear observation of the light field using pairs of opposite DM probing commands and then formulates the wavefront sensing as a least-squares problem. Building on that architecture, several improved wavefront sensing approaches have also been proposed, such as the Kalman filter method\cite{groff2013kalman} which combines information from previous AO control steps and the extended Kalman filter\cite{riggs2016recursive, pogorelyuk2019dark} which enables simultaneous incoherent source estimation. These improvements focus only on the formulation of the statistical estimation problem, not the DM probing and  image acquisition process itself. However, in space-based AO, the latter  is equally important. Unlike similar phase retrieval problems in other fields, such as quantitative phase imaging\cite{kellman2019physics}, wavefront sensing in space-based AO does not provide any science results in and of itself; the retrieved light field is only used for wavefront correction. Nevertheless, the ultimate objective is to observe  the faint incoherent astronomical objects hidden below the residual light from these coherent wavefront aberrations. Reducing the time spent on wavefront sensing and image acquisition significantly increases the available time for science observations.

In this paper, we propose an improvement to the stochastic modeling of a space-based AO system and accordingly introduce efficient wavefront sensing policies, where optimal DM probing commands and camera exposure times are used. Simulation results using a vortex coronagraph system\cite{mawet2009vector} show that our new approach achieves almost the same accuracy of the field estimation with fewer images and much shorter exposures, thus significantly reducing the time spent on wavefront correction.

\section{Space-based AO} \label{sec:AO}
Adaptive Optics in a space telescope is used to correct the wavefront aberrations in \textcolor{black}{the telescope optics and} the coronagraph instrument. Figure~\ref{fig:AO} shows a representative space telescope system equipped with both AO and a coronagraph. A coronagraph\cite{kasdin2003extrasolar, guyon2003phase, mawet2009vector, zimmerman2016shaped, trauger2016hybrid} is a type of optical device designed for imaging faint companions around a star. In a coronagraph instrument, a series of amplitude and phase masks work together to block out the on-axis starlight but transmit the off-axis light sources,  ultimately creating a high-contrast observation region, or so-called dark hole, in the image plane. However, the coronagraph is sensitive to \textcolor{black}{complex wavefront aberrations (both amplitude and phase errors)} in the optical system. When the wavefront is aberrated by the lens/mirror surface roughness, misalignments, and thermal effects, the focal plane observations are contaminated with bright speckles and the contrast in the dark hole is significantly degraded. In that case, DMs in the AO system are needed to restore the designed high contrast. The DM's surface can be controlled by applying various voltages to the DM actuators\cite{bifano2011adaptive}. As mentioned in Sec.~\ref{sec:intro}, DMs first apply probing commands to sense the light field, then apply control commands to compensate for the wavefront aberrations. This process is iterated, between the sensing and control, to dig a final dark hole. In this section, we describe the current space-based AO process, including system modeling and the wavefront sensing control (WFSC) policies.

\subsection{System modeling} \label{subsec:model}
The light propagation through the coronagraph can be modeled as a linear operation. \textcolor{black}{Figure~\ref{fig:AO} shows a space-based AO system equipped with a vortex coronagraph\cite{mawet2009vector}, which consists of a binary (fully transmissive or fully opaque) pupil plane mask, a focal plane vector vortex coronagraph (a pure phase mask which introduces an azimuthal phase ramp) and a binary Lyot stop mask. It defines a general architecture of a coronagraph instrument, however, the pupil plane masks can also be replaced by apodizers and the focal plane mask can also be replaced by amplitude masks in other types of coronagraphs\cite{kasdin2003extrasolar, guyon2003phase, mawet2009vector, zimmerman2016shaped, trauger2016hybrid}.}  With the coronagraph's pupil mask, focal plane mask, and the Lyot stop respectively denoted as $M_P$, $M_F$ and $M_L$, the relationship between the pupil plane field, $E_p$, and focal plane field, $E_f$, is
\begin{equation} \label{eq:coronagraph}
\begin{split}
E_f = {\mathcal{C}}\{E_p\} &= {\mathcal{F}} \{ M_L \cdot {\mathcal{F}}^{-1} \{ M_F \cdot {\mathcal{F}} \{M_P \cdot E_{p}\} \} \}\\
&= {\mathcal{F}} \{ M_L\} * [M_F \cdot ({\mathcal{F}} \{M_P\} * {\mathcal{F}} \{E_{p}\})],
\end{split}
\end{equation}
where $\mathcal{F}$ and $\mathcal{F}^{-1}$ represent the two-dimensional Fourier transform and inverse Fourier transform, respectively, and $\mathcal{C}$ is the composite linear coronagraph operator. To keep the representation clean, here we neglect a constant coefficient related to the light wavelength $\lambda$.

Deformable mirrors introduce phase perturbations to the incident light field. Assuming the incident light field with \textcolor{black}{complex wavefront aberrations} is given by $E_{ab}$, the corrected pupil wavefront downstream of the adaptive optics is
\begin{equation} \label{eq:dm}
E_p = E_{ab} \exp(i \frac{4 \pi \phi}{\lambda}),
\end{equation}
where $\phi$ is the surface height of the DM, $\lambda$ is the light wavelength, and the constant is $4$ instead of $2$ because \textcolor{black}{the change in the optical path length is twice the mirror displacement.}
%\footnote{In this equation, we assume only one DM,  which is located at the conjugate plane of the coronagraph pupil, is used in the AO system. A real space-based AO typically have two or more DMs to correct both amplitude and phase wavefront aberrations. However, except for an extra nested phase perturbation and Fresnel propogation, the two DM and one DM cases are not fundamentally different. We use the one DM formula here to keep the equation concise.}
The DM surface height is a two-dimensional function of the actuators' voltage commands, $\phi = \phi(u)$. It can be approximated as a linear superposition of each actuator's influence on the mirror surface\cite{prada2017characterization},
\begin{equation} \label{eq:infFun}
\phi(u) = \sum_{q=1}^{N_{act}} u_q f_q,
\end{equation}
where $N_{act}$ is the number of actuators on the DMs, $u_q$ is the voltage command to the q-th actuator, and $f_q$ is its unit voltage response on the DM surface. The DM surface varies over time when the AO control loop is running. After $k$ wavefront sensing and control steps, the accumulated DM surface height becomes
\begin{equation}
\phi_k = \sum_{t=1}^{k} \Delta \phi_t = \sum_{t=1}^{k} \sum_{q=1}^{N_{act}} \Delta u_{q, t} f_q = \phi_{k-1} + \phi (\Delta u_k),
\end{equation} 
where $\Delta \phi_t$ and $\Delta u_{q, t}$ are respectively the surface change and incremental voltage command at time step $t$, and $\Delta u_k$ is collection of all the actuators' voltage changes at step $k$.

Combining the Fourier optics modeling of the adaptive optics and the coronagraph (Eq.~\ref{eq:coronagraph}, Eq~\ref{eq:dm} and Eq.~\ref{eq:infFun}), the focal plane field after $k$ control steps is
\begin{equation} \label{eq:ssm}
E_{f, k} = \mathcal{C}\{E_{p, k}\} = \mathcal{C}\{E_{ab} \exp(i \frac{4 \pi \phi_k}{\lambda})\} = \mathcal{C}\{E_{p, k-1} \exp(i \frac{4 \pi \phi (\Delta u_k)}{\lambda})\}.
\end{equation}
Assuming the DM surface change is very small in each sensing and control step, Eq.~\ref{eq:ssm} can be linearized using a Taylor expansion. The space-based AO with coronagraph can thus be mathematically described as a linear time-varying (LTV) system,
\begin{equation} \label{eq:ssm2}
E_{f, k} \approx \mathcal{C}\{E_{p, k-1}\} + \mathcal{C}\{E_{p, k-1} i \frac{4 \pi \phi (\Delta u_k)}{\lambda}\} = E_{f, k-1} + G_{k-1} \Delta u_k,
\end{equation}
where $G_{k-1}$ is a linear projection modeling the DM's influence on the focal plane light field, which is also known as Jacobian matrix after $E_{f, k}$ and $\Delta u_k$ are discretized and vectorized.

The observations of the focal plane light field are the camera images perturbed by the DM probing commands. Letting the DM probing command be $\Delta u_k^p$, the corresponding camera image is
\begin{equation} \label{eq:observ}
I_{f, k}^p =  |E_{f, k} + G_{k} \Delta u_k^p|^2 + I_{in, k}.
\end{equation}
where $I_{in, k}$ is the incoherent signal not influenced by the DMs, such as the exoplanet light, and $|\cdot|^2$ represents the element-wise square of the amplitude of a complex vector/matrix. Equation~\ref{eq:ssm2} and Eq.~\ref{eq:observ} together describe a state space model (SSM) of the AO system. \textcolor{black}{Only monochromatic light is considered in the above optical modeling. However, it is straightforward to extend the above mathematical formula to the broadband case by defining each wavelength's SSM independently and then concatenating the state vectors of different wavelengths to formulate a broadband SSM.\cite{groff2015methods}}

\subsection{Wavefront sensing and control policies} \label{subsec:wfsc}
Wavefront control typically minimizes the total energy in the focal plane observation regions (dark holes) for the DM voltage commands according to Eq.~\ref{eq:ssm2}.  This control policy is usually referred to as electric field conjugation (EFC)\cite{borde2006high, give2007broadband} and can be formulated as a regularized quadratic programming problem,
\begin{equation}
\Delta u_k^{\star} = \text{arg} \min_{\Delta u_k} \| E_{f, k-1} + G_{k-1} \Delta u_k\|_2^2 + \alpha_k \|\Delta u_k\|_2^2,
\end{equation}
where $\|\cdot\|_2$ is the L-2 norm of a vector/matrix, $E_{f, k-1}$ is the discretized focal plane electric field in the dark holes,  $\Delta u_k^{\star}$ is the optimal DM control command and $\alpha_k$ is a Tikhonov regularizer. The Tikhonov regularizer is introduced to avoid unreasonably large commands exceeding the operation limit, since the equation of the electric field is an under-determined system (the number of actuators is smaller than the number of pixels in the search region).

Wavefront sensing solves the dual problem of estimating the focal plane light field. As  mentioned in Sec.~\ref{sec:intro}, the current benchmark wavefront sensing approach is the pair-wise DM probing and least-squares estimation\cite{borde2006high, give2011pair}. According to the observation model in Eq.~\ref{eq:observ},  differencing the perturbed images from opposite DM probing commands, $\pm \Delta u_k^p$, constructs a linear observation of the electric field,
\begin{equation} \label{eq:Idiff}
\begin{split}
I_{f, k}^{p, \pm} &= |E_{f, k} \pm p_k|^2 + I_{in, k} \\
\Delta I_{f, k}^p &=  I_{f, k}^{p, +} - I_{f, k}^{p, -} = 4 \Re\{p_k^{\dag} \circ E_{f, k}\},
\end{split}
\end{equation}
where $\dag$ represents the complex conjugate, $\circ$ represents the Hadamard product, and $p_k = G_{k} \Delta u_k^p$ is the focal plane electric field perturbation introduced by the probing commands. The estimation problem based on the linear observation can be thus formulated as a least-squares problem,
\begin{equation} \label{eq:bpe}
\hat{E}_{f, k} = \text{arg} \min_{E_{f, k}} \sum_{j=1}^{N_p} \|\Delta I_{f, k}^{p, j} - 4 \Re\{p_k^{j \dag} \circ E_{f, k}\}\|_2^2,
\end{equation}
where $\hat{E}_{f, k}$ is the estimated focal plane light field, $N_p$ is the number of pairs of opposite DM probing commands applied, $j$ is the index of the pair-wise probing commands, $\{p_k^{j}\}$ are the perturbations generated by $\{\Delta u_k^{p, j}\}$, and $\{\Delta I_{f, k}^{p, j}\}$ are the corresponding camera measurements.

\begin{figure}[h!]
\centering\includegraphics[width=6cm]{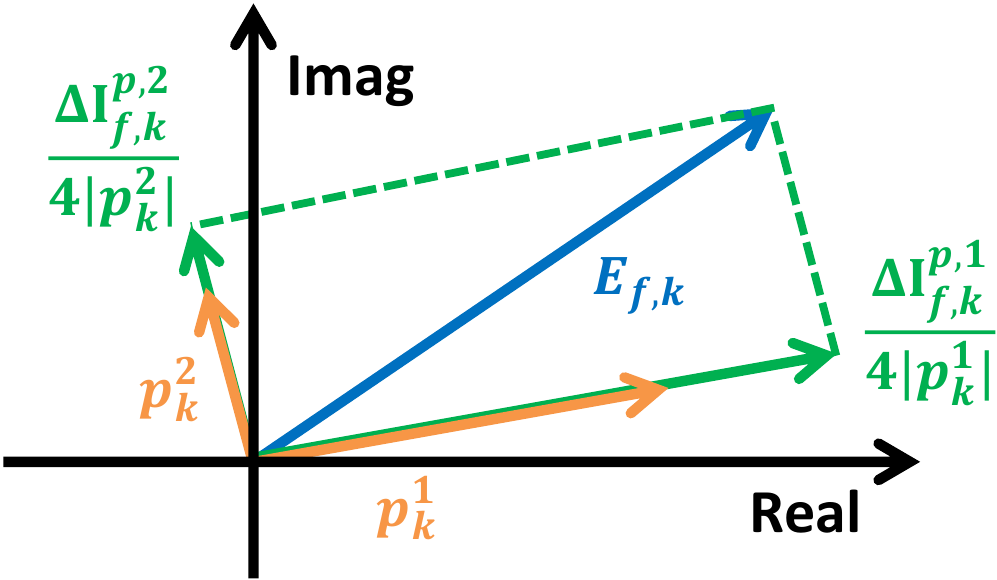}
\caption{A graphical interpretation of pair-wise probes and difference images.}
\label{fig:pairwise}
\end{figure}

By visualizing the pair-wise probing observation in a complex plane, as shown in Fig.~\ref{fig:pairwise}, we can see that the difference images measure the projections of the complex electric field on the perturbation directions. At least 2 pairs of probes are needed to estimate the field; the probing directions should be as different as possible. In addition, the probes should  modulate the whole observation area, otherwise the difference between the positive and negative images at some pixels would be too small to be used for regression. To satisfy the above requirements, currently the most popular DM probing policy is to generate two-dimensional sinc waves on the DM surface\cite{groff2015methods},
\begin{equation} \label{eq:sinc}
\begin{split}
\Delta \phi_k^{p, j} = \phi(\Delta u_k^{p, j}) &= \beta_k \text{sinc}(ax) \ \text{sinc}(by) \ \cos(cx+\psi_j) \\
&\text{or} \ \beta_k \text{sinc}(ax) \ \text{sinc}(by) \ \cos(cy+\psi_j),
\end{split}
\end{equation}
where $(x, y)$ are the coordinates in the pupil plane and $\beta_k, a, b, c$ and $\psi_j$ are constants. The Fourier transform of the above DM surface shape (assuming the first formula) is
\begin{equation} \label{eq:rect}
\begin{split}
\mathcal{F}\{\Delta \phi_k^{p, j}\} &= \frac{1}{2} \beta_k \exp(i\psi_j) \ \text{rect}(a x^{\prime}, b y^{\prime}) * \delta(x^{\prime}-c) \\
&+ \frac{1}{2} \beta_k \exp(-i\psi_j) \ \text{rect}(a x^{\prime}, b y^{\prime}) * \delta(x^{\prime}+c)
\end{split}
\end{equation}
which produces two symmetric uniform rectangles with opposite phase diversities in the Fourier domain. According to Eq.~\ref{eq:coronagraph}, the coronagraph operator is similar to the Fourier transform except for some extra kernel convolutions, so the focal plane perturbations created by the sinc probes should still have a relatively uniform and symmetric structure \textcolor{black}{(see Fig.~\ref{fig:empSinc} and case (a) in Sec.~\ref{sec:simulation})}. The phase (or the direction in the complex plane) of each probe can be adjusted by changing the shift term $\psi_j$. Typically, a set of probing phases that uniformly cover the range of $[0, \pi]$ are selected. The amplitude of the sinc waves $\beta_k$ are computed based on the mean probing contrast, that is, the probe amplitude in units of contrast, $C_k^p = \sum |p_k|^2 / N_{pix}$, where $N_{pix}$ is the number of pixels in the observation regions. A widely used heuristic law for determining the probing contrast\cite{riggs2018fast} is 
\begin{equation} \label{eq:probeContrast}
C_k^p = \min(\sqrt{10^{-5}\times C_k}, 10^{-4}),
\end{equation}
where $C_k = \sum|E_{f, k}|^2 / N_{pix}$ is the mean contrast of the original observation region. By using Eq.~\ref{eq:probeContrast}, the probe contrast is neither too large nor too small, which gives a relatively accurate measurement of the original focal plane electric field.

Two more advanced wavefront sensing approaches are the Kalman Filter and the (iterated) extended Kalman filter. The Kalman filter (KF) \cite{groff2013kalman} incorporates the state transition information and solves a weighted least-squares problem for the electric field estimation,
\begin{equation} \label{eq:kf}
\hat{E}_{f, k} = \text{arg} \min_{E_{f, k}} \sum_{j=1}^{N_p} \|\Delta I_{f, k}^{p, j} - 4 \Re\{p_k^{j \dag} \circ E_{f, k}\}\|_{R_k}^2 + \|E_{f, k} - E_{f, k-1} - G_{k-1} \Delta u_k\|_{P_k}^2,
\end{equation}
where $\|\cdot\|_{R_k}$ and $\|\cdot\|_{P_k}$ are the matrix weighted norms that balance the importance of the information from current observations and state transitions. \textcolor{black}{The matrix weights $R_k$ and $P_k$ are computed based on the level of process noises and observation noises of the system.} The Extended Kalman Filter (EKF) \cite{riggs2016recursive, pogorelyuk2019dark} directly solves a phase-retrieval-like non-convex inverse problem,
\begin{equation} \label{eq:ekf}
\begin{split}
\hat{E}_{f, k}, \hat{I}_{in, k} = \text{arg} \min_{E_{f, k}, I_{in, k}} &\sum_{j=1}^{N_p} \|I_{f, k}^{p, j} - |E_{f, k} + p_k^j|^2 - I_{in, k}\|_{R_k}^2 \\
&+ \|E_{f, k} - E_{f, k-1} - G_{k-1} \Delta u_k\|_{P_k}^2 + \|I_{in, k} - I_{in, k-1}\|_{P_k^{\prime}}^2,
\end{split}
\end{equation}
where now $N_p$ is the number of DM probing commands instead of pairs of probing commands. These two approaches improve the formulation of the statistical estimation problem by better utilizing the information acquired. However, the DM probing and image acquisition policies stay the same as in the benchmark case.

\section{Optimal probing policy and efficient camera integration time} \label{sec:optimal}
Based on the space-based AO framework described in Sec.~\ref{sec:AO}, we now present our new approach for optimizing the DM probing policies and camera integration times. We here mainly consider the case of pair-wise probing and estimation.
%We first propose a new noise model of the system, and then explain how the optimal DM probes and camera exposure depend on the system noise coefficients.
\subsection{A stochastic model of system noises} \label{subsec:noiseModel}
The state-space model introduced in Sec.~\ref{subsec:model} is  a stochastic process with additive process noise, $w_k$, in the state transition equation and observation noise, $n_k$, in the state observation equation,
\begin{equation} \label{eq:noisySSM}
\begin{split}
E_{f, k} &= E_{f, k-1} + G_{k-1} \Delta u_k + w_k,\\
I_{f, k}^p &=  |E_{f, k} + p_k|^2 + I_{in, k} + n_k.
\end{split}
\end{equation}
Although not explained in detail in Sec.~\ref{subsec:wfsc}, the weighting matrices $R_k$, $P_k$, and $P_k^{\prime}$ in the Kalman filter and the extended Kalman filter depend on the statistics of these model noises. 

Based on the experimental observations, we found that the process noises could be modeled as zero-mean Gaussian,
\begin{equation} \label{eq:processNoise}
[\Re\{w_k\}, \Im\{w_k\}] \sim \mathcal{N}(0, \sigma_k^2 \mathbb{I} ),
\end{equation}
where $\mathbb{I}$ is the identity matrix and the variance is approximately proportional to the summation of a constant and the electric field change multiplied by another constant,
\begin{equation} \label{eq:processCov}
\sigma_k^2 = q_0 + q_1|\Delta E_{f, k}|^2 = q_0 + q_1|G_{k-1} \Delta u_k|^2.
\end{equation}
The observation equation and observation noise can thus be distributed using Eq.~\ref{eq:processCov}, 
\begin{equation} \label{eq:observNoise}
\begin{split}
I_{f, k}^p &=  |E_{f, k} + p_k + w_k^p|^2 + I_{in, k} + n_k^d\\
& =  |E_{f, k} + p_k|^2 + I_{in, k} + n_k^d+ 2|(E_{f, k} + p_k)^{\dag} \circ w_k^p| + |w_k^p|^2  \\
\end{split}
\end{equation}
where $w_k^p$ is the DM probing noise and $n_k^d$ is the detector noise. The observation noise can be modeled as a non-zero mean Gaussian,
\begin{equation} \label{eq:observNoiseStats}
\begin{split}
n_k &=  n_k^d + 2|(E_{f, k} + p_k)^{\dag} \circ w_k^p| + |w_k^p|^2 \sim \mathcal{N}(\mu_k, \nu_k^2 \mathbb{I}),\\
\mu_k &= 2 (q_0 + q_1 |p_k|^2),\\
\nu_k^2 &=  r_0 + r_1 \frac{\sum (|E_{f, k} + p_k|^2 + I_{in, k})}{N_{pix}} + \frac{\sum 4|E_{f, k} + p_k|^2}{N_{pix}}(q_0 + q_1 |p_k|^2),
\end{split}
\end{equation}
of which the covariance consists of three parts: the first term from the camera readout noise is a fixed constant, the second term from the Poisson noise is proportional to the image intensity, and the third term from DM probing has a similar formula to Eq.~\ref{eq:processCov}. The readout noise and Poisson noise terms are related to the camera integration time, $t$, via
\begin{equation} \label{eq:observNoiseCoef}
r_0 = \frac{n_r^2}{(\text{flux} \cdot t)^2} = \frac{\bar{r}_0}{t^2}, \quad r_1 = \frac{1}{\text{flux}\cdot t} = \frac{\bar{r}_1}{t},
\end{equation}
where \textcolor{black}{the flux (proportional to the brightness of the planet's host star) is the number of photons hitting the detector in unit time at the center pixel of the starlight point spread function (PSF)}, $n_r$ is the camera's readout standard deviation, and \textcolor{black}{$\bar{r}_0$ and $\bar{r}_1$} are the normalized noise coefficients. The covariance of the pair-wise probing observation is thus
\begin{equation} \label{eq:pairwiseNoise}
\nu_{k, \text{pair}}^2 = \nu_{k, +}^2 + \nu_{k, -}^2 = 2 r_0 + 2 (r_1 + 4 q_0) (C_k + C_k^p + C_k^{in})+ 8 q_1 C_k^p (C_k + C_k^p),
\end{equation}
where \textcolor{black}{$C_k=\sum |E_{f, k}|^2 / N_{pix}$ and $C_k^p=\sum |p_k|^2 / N_{pix}$ are respectively the mean coherent contrast and mean probing contrast} defined in Sec.~\ref{subsec:wfsc} and $C_k^{in} = \sum I_{in, k} / N_{pix}$ is the mean incoherent contrast.

\subsection{Optimal probing contrast} \label{subsec:optContrast}
The optimal probing contrast should minimize the covariances of the observation noises. According to Fig.~\ref{fig:pairwise}, the projection measurement of the electric field is $z_k = \Delta I_{f, k}^p/(4 |p_k|)$, so the corresponding observation noise variance is
\begin{equation}
\text{Var}(z_k) = \frac{\nu_{k, \text{pair}}^2}{16|p_k|^2} = \frac{r_0 + (r_1 + 4 q_0) (C_k + C_k^{in})}{8 C_k^p} + \frac{q_1 C_k^p}{2} + \frac{r_1}{8} + \frac{q_0 +q_1 C_k}{2}.
\end{equation}
According to the inequality of arithmetric and geometric means (AM-GM inequality)\cite{hirschhorn2007gm},
\begin{equation} \label{eq:optimalProbeContrast}
\begin{split}
&\text{Var}(z_k) \geq \frac{\sqrt{q_1 [r_0 + (r_1 + 4 q_0) (C_k + C_k^{in})]}}{2} + \frac{r_1}{8} + \frac{q_0 +q_1 C_k}{2},\\
&\text{with equality if and only if } C_k^p = \sqrt{\frac{r_0}{4 q_1} + \frac{r_1 + 4 q_0}{4 q_1} C_k^{in} + \frac{r_1 + 4 q_0}{4 q_1} C_k}.
\end{split}
\end{equation}
We now have a theoretical solution for the optimal probing contrast. When the camera readout noise is very small ($r_0 \rightarrow 0$) and the incoherent contrast is small ($C_k^{in} \rightarrow 0$), this theoretical solution is similar to the heuristic law in Eq.~\ref{eq:probeContrast}, \textcolor{black}{where the mean probing contrast is proportional to the square root of the mean coherent field contrast}. However, our optimal law has a clearer physical meaning and varies for different systems, depending on the DMs, the coronagraph, and the detectors used.

\subsection{Optimal probing shape}
%The optimal probing contrast only considers the variance of a single projection measurement. However, the probing shape also plays an important role in the state estimation accuracy. 
With $n$ pairs of probes and difference images, $\{p_k^j, \Delta I_{f, k}^{p, j}\}, j = 1, \cdots, N_p$, we can write  the over-determined linear observation equation of the electric field,
\begin{equation}
\tilde{z}_k = 
\begin{bmatrix}
z_k^1\\
\vdots \\
z_k^{N_p}
\end{bmatrix}
=
\begin{bmatrix}
\Delta I_{f, k}^{p, 1} / (4 |p_k^1|)\\
\vdots \\
\Delta I_{f, k}^{p, N_p} / (4 |p_k^{N_p}|)
\end{bmatrix}
=
\begin{bmatrix}
\cos(\theta_{k, 1}) & \sin(\theta_{k, 1})\\
\vdots & \vdots \\
\cos(\theta_{k, N_p}) & \sin(\theta_{k, N_p})
\end{bmatrix}
\begin{bmatrix}
\Re\{E_{f, k}\} \\
\Im\{E_{f, k}\}
\end{bmatrix}
\overset{\Delta}{=} H_k x_k,
\end{equation}
where $\theta_{k, j}=\text{arctan2}(\Re\{p_k^j\}, \Im\{p_k^j\})$ defines the orientation of the complex probe perturbation, $p_k^j$. The estimated mean and covariance of the electric field are
\begin{equation}
\begin{split}
&\hat{x}_k = (H_k^T H_k)^{-1} H_k^T \tilde{z}_k,\\
&\text{Cov} (\hat{x}_k) = (H_k^T H_k)^{-1} H_k^T
\begin{bmatrix}
\text{Var}(z_k^{1}) & & \\
& \ddots & \\
& & \text{Var}(z_k^{N_p})
\end{bmatrix}
 H_k^{-T} (H_k^T H_k)^{-1},
\end{split}
\end{equation}
\textcolor{black}{where $\text{Cov} (\hat{x}_k)$, consisting of $H_k$ and $\{\text{Var}(z_k^j)\}$, are functions of $\{\Delta u_k^{p, j}\}$ and optical model parameters (Jacobian matrix, $G_k$, and noise coefficients, $q_0, q_1, r_0, r_1$, defined in Sec.~\ref{subsec:noiseModel}). The optimal probe shape can thus be computed by minimizing the log determinant of this covariance matrix with respect to the DM probing voltage commands},
\begin{equation} \label{eq:varmin}
\{\Delta u_k^{p, j}\} =  \text{arg} \min_{\{\Delta u_k^{p, j}\}} \log |\text{Cov} (\hat{x}_k)| + \mathcal{P}(\{\Delta u_k^{p, j}\}),
\end{equation}
where $\mathcal{P}(\cdot)$ is a user-chosen regularizer that prevents ill-posed solutions. One useful choice of the regularizer is a Tikhonov regularization of the DM probing voltage commands. This policy is typically called variance-minimizing \cite{powell2012optimal} in active learning and optimal experiment design.

When we have only two pairs of probing commands, the log determinant of the estimation covariance can be simplified to an easily interpreted formula,
\begin{equation}
\log |\text{Cov} (\hat{x}_k)| = \log \text{Var}(z_k^1) + \log \text{Var}(z_k^2) - 2 \log |\sin(\theta_{k, 2} - \theta_{k, 1})|.
\end{equation} 
Minimizing the first two terms makes the DM probing commands satisfy the optimal probing contrast criterion in Sec.~\ref{subsec:optContrast}, while minimizing the third term makes the complex perturbations as perpendicular as possible to each other.

\subsection{Efficient camera integration time} \label{subsec:optTime}
Assuming the optimal probing policy is applied, we can now determine the best camera integration time by analyzing the signal-to-noise-ratio (SNR). The SNR of the camera image is defined as
\begin{equation} \label{eq:snr}
\begin{split}
\frac{1}{\text{SNR}} = \frac{\text{Var}(z_k)}{C_k} &= \frac{1}{2} \sqrt{q_1 [\frac{r_0}{C_k^2} + \frac{r_1}{C_k} + \frac{r_1 C_k^{in}}{C_k^2} + \frac{4 q_0 (C_k + C_k^{in})}{C_k^2}]} + \frac{r_1}{8 C_k} + \frac{q_0 +q_1 C_k}{2 C_k}\\
&= \frac{1}{2} \sqrt{q_1 [\frac{\bar{r}_0}{C_k^2t^2} + \frac{\bar{r}_1}{C_k t} + \frac{\bar{r}_1 C_k^{in}}{C_k^2 t} + \frac{4 q_0 (C_k + C_k^{in})}{C_k^2}]} + \frac{\bar{r}_1}{8 C_k t} + \frac{q_0 +q_1 C_k}{2 C_k}\\
&\overset{\Delta}{=} \frac{1}{2} \sqrt{q_1 [\frac{\bar{r}_0}{\gamma^2} + \frac{\bar{r}_1}{\gamma} + \frac{\bar{r}_1 C_k^{in}}{\gamma C_k} + \frac{4 q_0 (C_k + C_k^{in})}{C_k^2}]} + \frac{\bar{r}_1}{8 \gamma} + \frac{q_0 +q_1 C_k}{2 C_k},
\end{split}
\end{equation}
\textcolor{black}{where $\gamma \overset{\Delta}{=} C_k t$. The SNR is maximized when $\gamma \to +\infty$, i.e. the camera integration time becomes infinitely large.} However, the last term in the square root and the last term of the equation are not influenced by the integration time, so the marginal benefit becomes very small when the integration time exceeds a certain threshold.

Based on this observation, we can define an adaptive camera exposure policy, $t_k = \max(\gamma / C_k, t_{min})$, which results in shorter integration times when the contrast is low but longer for high contrast. We set a minimum integration time, $t_{min}$, to avoid too short an integration time which would result in  abnormal detector effects and too large a probing contrast that exceeds the DM linear operation regime. This policy results in a simplification of Eq.~\ref{eq:snr},
\begin{equation} \label{eq:adaptiveTime}
\frac{1}{\text{SNR}} \approx \frac{1}{2} \sqrt{q_1 [\frac{\bar{r}_0}{\gamma^2} + \frac{\bar{r}_1}{\gamma} + \frac{4 q_0}{C_k}]} + \frac{\bar{r}_1}{8 \gamma} + \frac{q_0 +q_1 C_k}{2 C_k},
\end{equation} 
when $C_k \gg C_k^{in}$. Given any SNR, we can solve for the camera integration time according to the above equation.

\section{Numerical experiments} \label{sec:simulation}
\subsection{Simulation setup} \label{subsec:setup}
In this section, we show the results of a simulation of a space-based AO system to demonstrate our new optimal probing and camera exposure policies. The layout of the system is almost identical to Fig.~\ref{fig:AO} except for an additional DM. A simple vortex coronagraph is used, consisting of a circular pupil aperture, a charge six vortex phase mask, and a Lyot stop. Two DMs with $34 \times 34$ actuators are used in the AO system, with the first placed at the conjugate plane of the coronagraph pupil plane. The image plane observation region (dark hole) is an annular area extending from $3 - 9 \lambda/D$, where $\lambda=635nm$ is the wavelength of the starlight and $D=1cm$ is the diameter of the coronagraph pupil mask. \textcolor{black}{Both amplitude and phase wavefront aberrations are introduced in our simulation.} In the wavefront sensing and control loop, both DMs are used to correct the wavefront aberrations but only the first one is used for probing the dark hole field. The system parameters defined in Sec.~\ref{sec:optimal} are listed in Tab.~\ref{tab:params}. These parameters could either be easily measured (such as the flux and the detector statistics, $n_r$) or be computed using system identification algorithms\cite{sun2018identification, sun2018neural} (such as the process noise parameters $q_0$ and $q_1$).
\begin{table}[]
\centering
\begin{tabular}{|c | c | c | c | c | c |} 
\hline
\text{flux} & $n_r$ & $q_0$ & $q_1$ & $\bar{r}_0$ & $\bar{r}_1$ \\
\hline
$2 \times 10^9$ & 12 & $10^{-14}$ & $0.05$ & $3.6 \times 10^{-17}$ & $5 \times 10^{-10}$ \\ 
\hline
\end{tabular}
\caption{AO system parameters in the numerical simulations.}
\label{tab:params}
\end{table}

We  explore five wavefront sensing approaches: (a) sinc wave probes (Eq.~\ref{eq:sinc}) and the heuristic probing contrast (Eq.~\ref{eq:probeContrast}), (b) sinc probes with optimal probing contrast (Eq.~\ref{eq:optimalProbeContrast}), (c) optimized probes initialized with sinc waves (Eq.~\ref{eq:varmin}), (d) optimized probes randomly initialized, and (e) the probing policy in (c) with adaptive camera integration times (Eq.~\ref{eq:adaptiveTime}). The camera integration time for each image in the first four cases is fixed at one second. We choose that number so that the WFSC  reaches the desired high contrast (around $5 \times 10^{-10}$) for the simulated flux. In space, the flux is very low, so the camera exposure times in a real space mission would be much longer. Our simulations only reflect the relative time needed for different approaches. Although the DM probing and camera exposure policies are different, an identical least-squares estimator (Eq.~\ref{eq:bpe}) is applied for all  cases.

\subsection{Results} \label{subsec:discuss}
\begin{figure}[h!]
\centering\includegraphics[width=14cm]{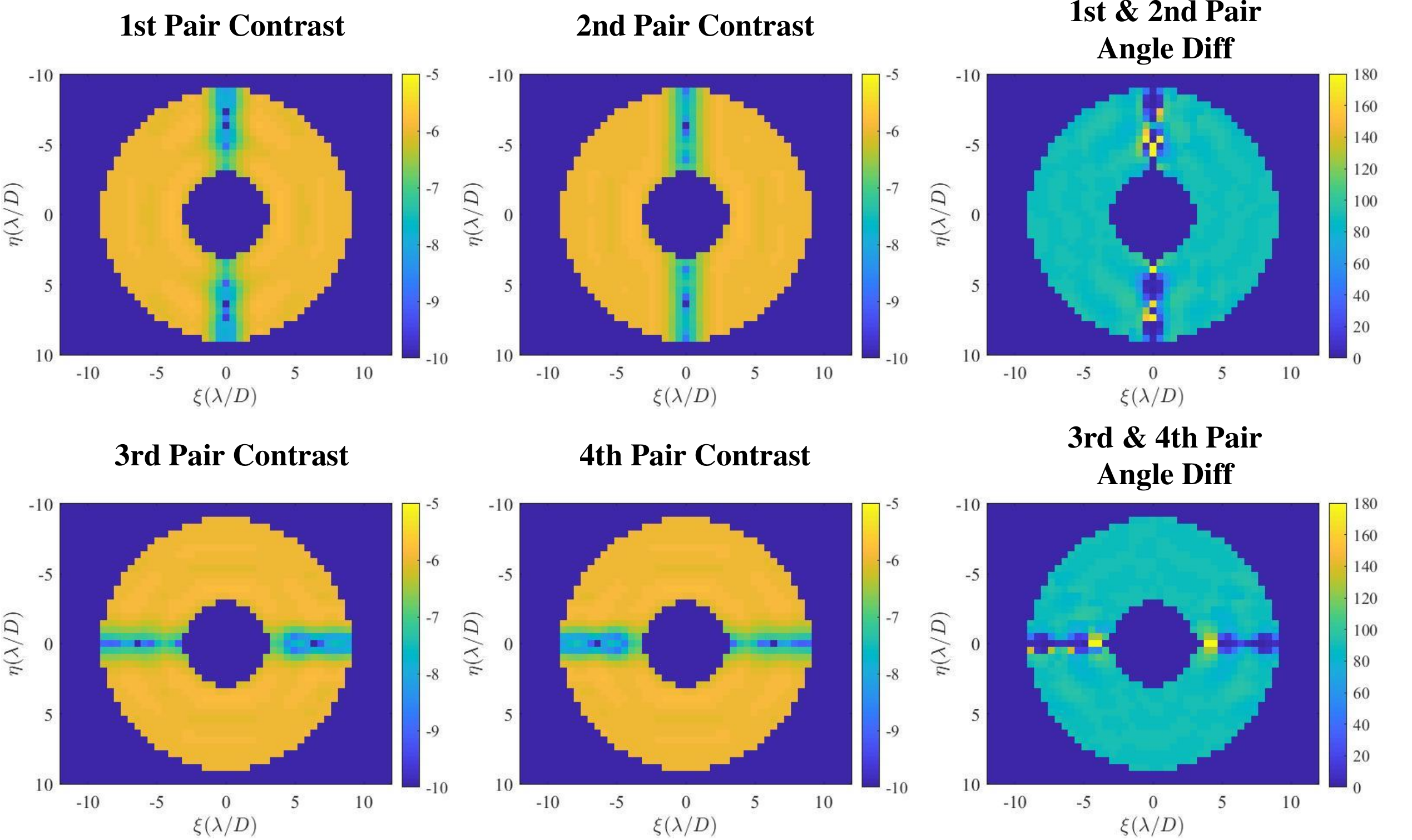}
\caption{Focal plane perturbations at the first control step caused by the empirical sinc waves with optimal amplitudes. $(\xi, \eta)$ are the focal plane coordinates, and $\lambda/D$ is the unit of the field of view (FOV), where $\lambda$ is the star light wavelength and $D$ is the telescope aperture size. The first two columns and the third column respectively display the probes' contrasts in log scale, $\{|p_k^{1}|^2,\cdots, |p_k^{4}|^2\}$, and the angle differences, $\{|\angle{p_k^1} - \angle{p_k^2}|, |\angle{p_k^3} - \angle{p_k^4}|\}$.}
\label{fig:empSinc}
\end{figure}

\begin{figure}[h!]
\centering\includegraphics[width=8cm]{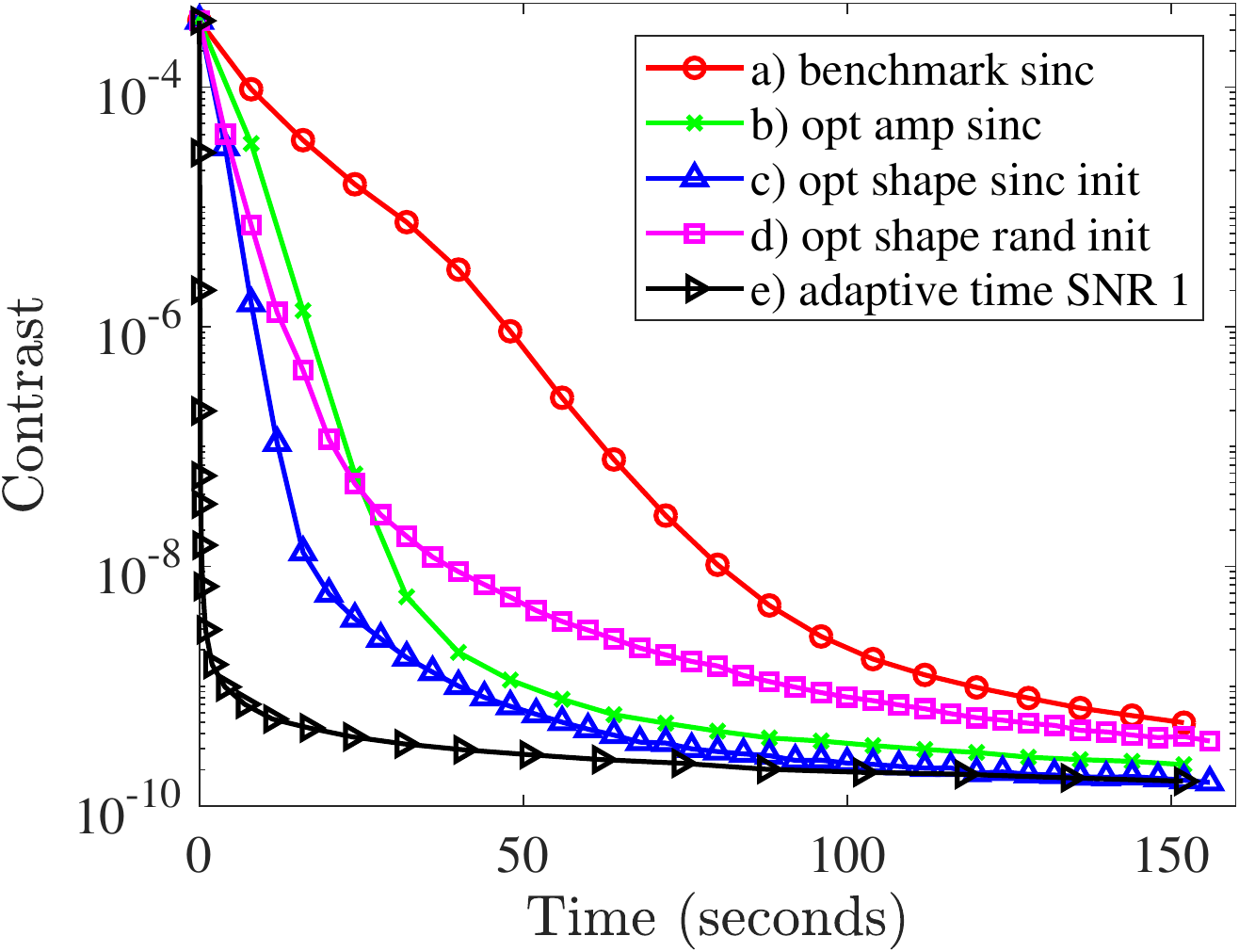}
\caption{WFSC simulation contrast curves with different DM probing and camera exposure policies, (a) benchmark case using sinc wave probes and the heuristic probing contrast, (b) sinc probes with optimal amplitude, (c) optimized probes initialized with sinc waves, (d) optimized probes randomly initialized, and (e) adaptive camera exposure with probing policy in (c).}
\label{fig:contrastCurve}
\end{figure}

Since the annular dark hole setup is used in our simulation, cases (a) and (b) need to at least apply four pairs of sinc waves for probing. Only two pairs of sinc probes are not enough. As shown in Fig.~\ref{fig:empSinc}, either the pixels located on the x-axis or on the y-axis are not well modulated by two pairs of sinc probes (probe intensity is too low), so we have to switch the probing axis to fully cover the dark hole regions.
Typically, the first and the second pair perturb two symmetric regions on the left and the right, and the third and the fourth pair perturb two regions on the top and the bottom. The phase shifts between the first two pairs and the second two pairs are both 90 degrees, which results in almost orthogonal electric field perturbations in the focal plane. As indicated in Fig.~\ref{fig:contrastCurve}, where we show the changes of image contrast over time in wavefront control, even though both cases use sinc probing profiles, case (b) with optimal amplitudes uses a much shorter time (fewer images) than the benchmark case (a). The optimal probing contrast law guides us to collect images with smaller observation noises.

\begin{figure}[h!]
\centering\includegraphics[width=14cm]{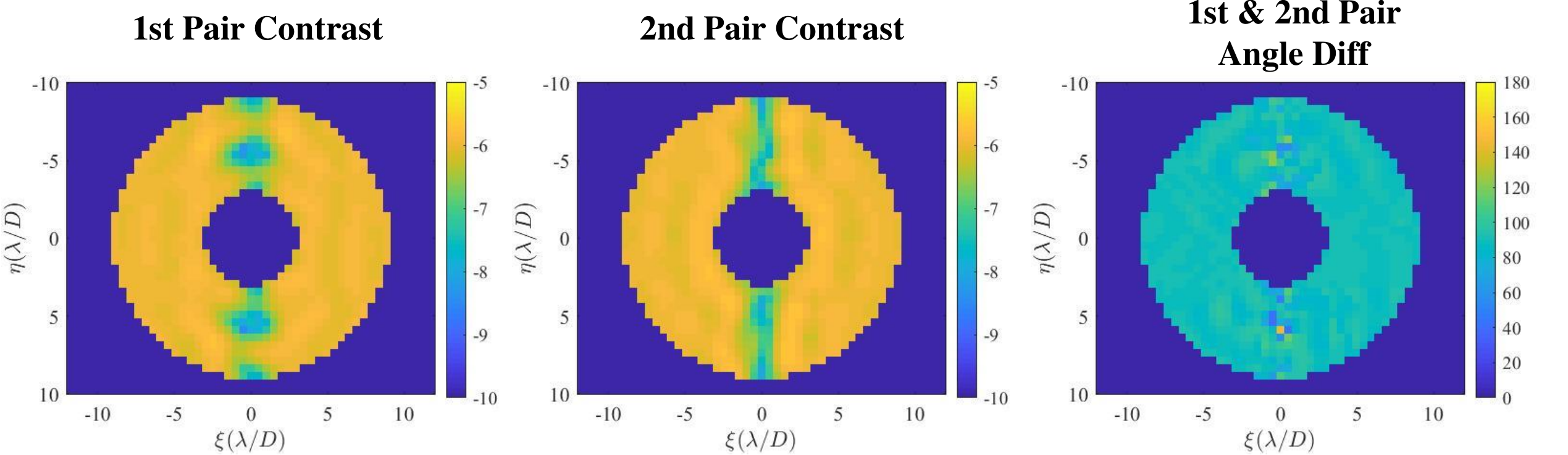}
\caption{Focal plane perturbations at the first control step caused by the optimized probe shapes with Sinc wave initialization. The first two columns and the third column respectively display the probes' contrasts in log scale, $\{|p_k^{1}|^2, |p_k^{2}|^2\}$, and the angle differences, $\{|\angle{p_k^1} - \angle{p_k^2}|\}$.}
\label{fig:optSinc}
\end{figure}

\begin{figure}[h!]
\centering\includegraphics[width=14cm]{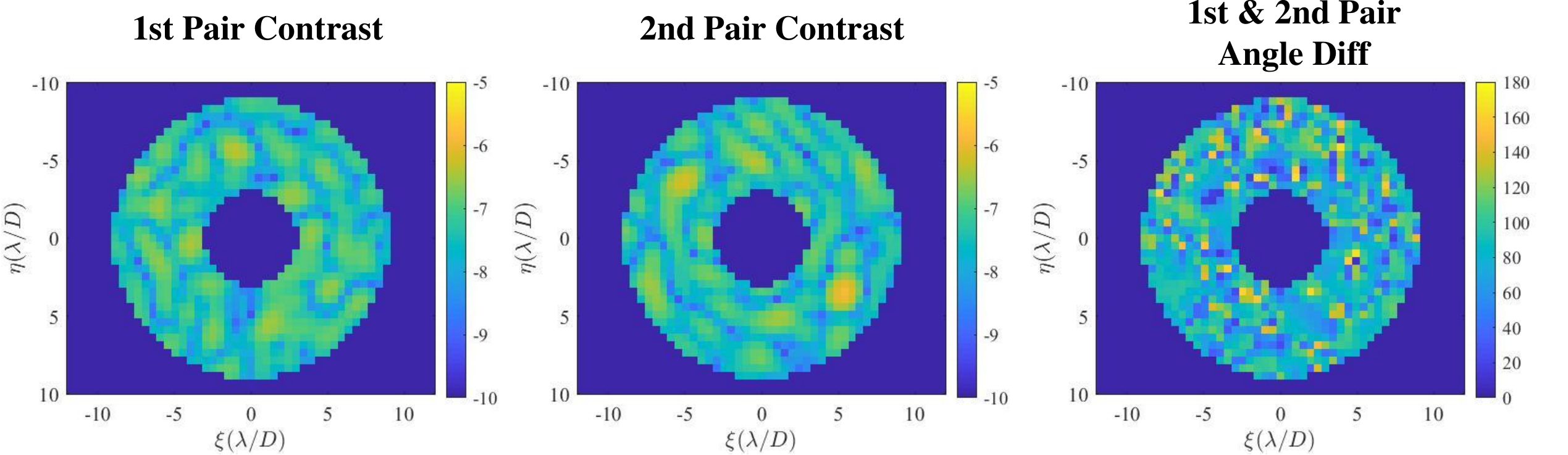}
\caption{Focal plane perturbations at the first control step caused by the optimized probe shapes with random initialization. The first two columns and the third column respectively display the probes' contrasts in log scale, $\{|p_k^{1}|^2, |p_k^{2}|^2\}$, and the angle differences, $\{|\angle{p_k^1} - \angle{p_k^2}|\}$.}
\label{fig:optRand}
\end{figure}

With the DM probing shapes optimized in cases (c) and (d), two-pairs-of-probe sensing becomes possible which further reduces the time and number of images needed for WFSC to achieve a high contrast. The probe shape optimization problem in Eq.~\ref{eq:varmin} is solved using an Adam optimizer\cite{kingma2014adam}. The initialization highly influences the final solutions since it is a non-convex program. As can be seen in Fig.~\ref{fig:contrastCurve}, case (c) with a sinc wave initialization performs better. It slightly modifies the sinc probe shapes and introduces electric field perturbations to the previously unmodulated axial regions. As indicated by Fig.~\ref{fig:optSinc}, the probe contrasts of the axial pixels now increase from below $10^{-9}$ to above $10^{-7}$, and the probe angle differences become almost $90^{\circ}$. However, with a random initialization, the focal plane perturbation is not always uniform (see Fig.~\ref{fig:optRand}) because the solution becomes easily stuck at a local minimum. Although it performs well at the beginning, case (d) doesn't beat case (b) in the later stage.

\begin{figure}[h!]
\centering\includegraphics[width=8cm]{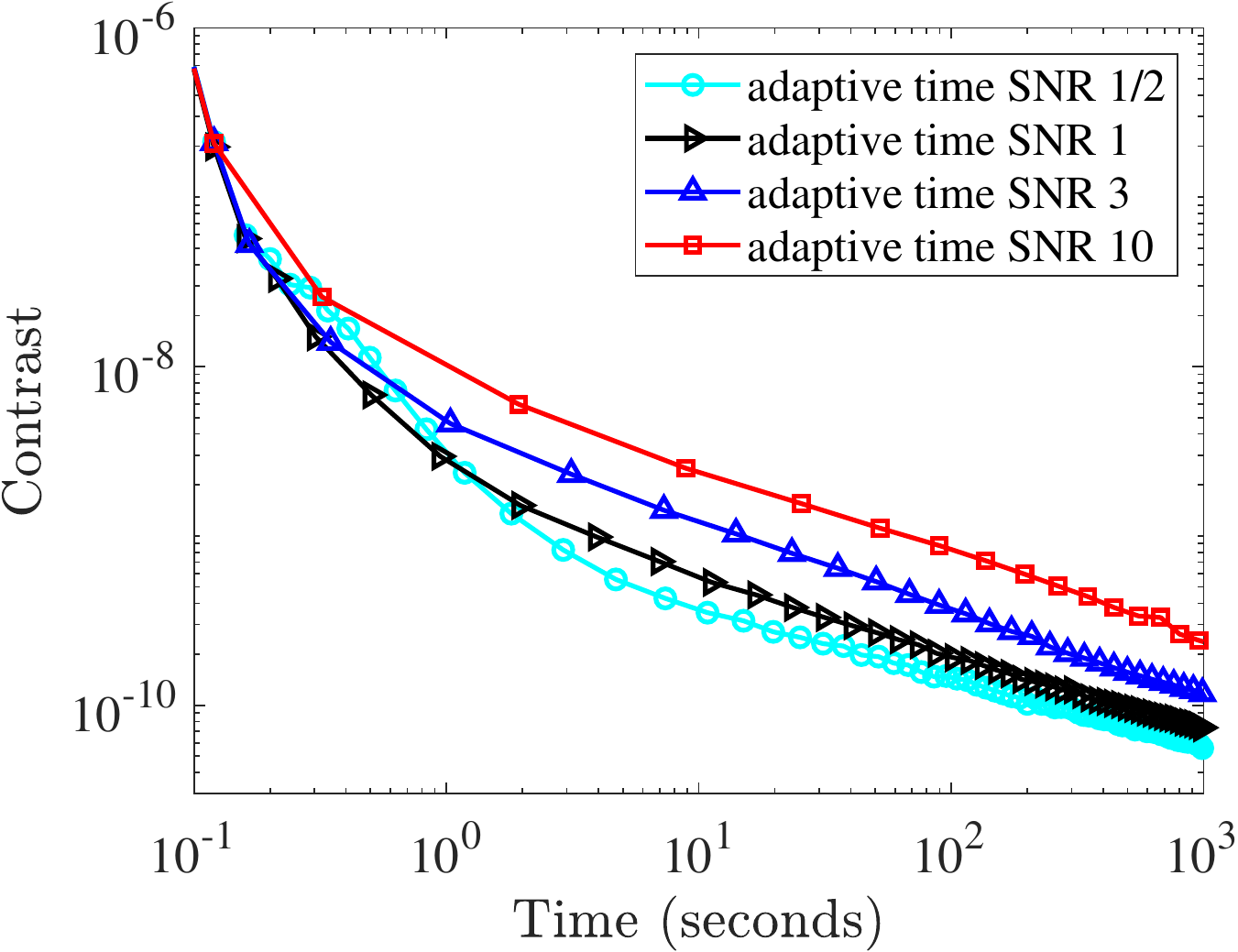}
\caption{WFSC contrast curves with optimal probing shapes and adaptive camera exposure policies defined at different SNRs, $1/2$, $1$, $3$ and $10$.}
\label{fig:adaptiveTime}
\end{figure}

Case (e), the adaptive exposure policy, is simulated using the same probing policy as in case (c) but with the adaptive integration time. The adaptive exposure policy significantly reduces the WFSC time while still reaching a high contrast, as shown in Fig.~\ref{fig:contrastCurve}, because short camera integration times are sufficient for the wavefront sensing at low contrast.  In the figure it appears that the adaptive camera exposure policy levels out at $3 \times 10^{-10}$ and converges to the fixed exposure case after one hundred seconds. However, this is not true and is purely because of the long camera exposure time at high contrast. In Fig.~\ref{fig:adaptiveTime}, we show a log-log contrast-versus-time graph from the adaptive camera exposure policies defined at four different SNRs. As can be seen, the contrast is still going down after a hundred seconds (can reach below $10^{-10}$), and the logarithm of the final contrast is inversely proportional to the logarithm of the WFSC control time used. In contrast, the WFSC contrast curves using fixed integration time do get stuck, because the collected high contrast probing images have very low SNRs. Typically, reducing the SNR speeds up the wavefront correction. However, when the SNR is below 1, such as in the case of $\text{SNR}=1/2$, WFSC is a little slower at the beginning because the estimation is not very robust. Even worse, when the SNR reaches below $1/2$ (not shown), the WFSC no longer works. That also explains why the fixed integration time policy get stuck after reaching a high contrast due to low-SNR images. Therefore, the adaptive camera exposure policy defined at $\text{SNR}=1$ is the best choice.

\section{Conclusion and future work}
In this paper, we have proposed a new stochastic model of a space-based AO system and developed optimal DM probing policies and adaptive camera exposure policies based on that. Our new approach enables  efficient  wavefront sensing, so the AO system can reach a high contrast within a much shorter time. We demonstrated the new approach by simulating a telescope system with a vortex coronagraph.

Future work includes the validation of the new methods in experiment. \textcolor{black}{We also want to explore the applications of our efficient wavefront sensing algorithms in correcting non-common-path errors in ground-based telescopes.} In addition, we also plan to investigate the optimal probing and camera exposure policies for nonlinear wavefront sensing algorithms, such as the extended Kalman filter in Eq.~\ref{eq:ekf}.

\acknowledgments 
This work was performed under contract to National Aeronautics and Space Administration (NASA), award number AWD1004730.

%%%%% References %%%%%

\bibliography{report}   % bibliography data in report.bib
\bibliographystyle{spiejour}   % makes bibtex use spiejour.bst

%%%%% Biographies of authors %%%%%

\vspace{2ex}\noindent\textbf{He Sun} is a postdctoral researcher at California Institute of Technology. He received his Ph.D from Princeton University in 2019 and his B.S. from Peking University in 2014. His research focuses on adaptive optics and computational imaging, which combines multidisciplinary ideas from optics, control, signal processing and machine learning. He is a member of the American Astronomical Society and the SPIE.

\vspace{2ex}\noindent\textbf{N. Jeremy Kasdin} is a Professor of Mechanical and Aerospace Engineering at Princeton University. He is the Principal Investigator of Princeton’s High Contrast Imaging Laboratory and Coronagraph Adjutant Scientist for WFIRST, the Wide Field InfraRed Survey Telescope. He received his Ph.D. from Stanford University in 1991. Professor Kasdin’s research interests include space systems design, space optics and exoplanet imaging, orbital mechanics, guidance and control of space vehicles, optimal estimation, and stochastic process modeling. He is an Associate Fellow of the American Institute of Aeronautics and Astronautics and member of the American Astronomical Society and the SPIE.

\vspace{2ex}\noindent\textbf{Robert Vanderbei} is a Professor of Operations Research and Financial Engineering at Princeton University. He also holds courtesy appointments in the Department of Mathematics, Astrophysics, Computer Science, and Mechanical and Aerospace Engineering. He received his Ph.D. from Cornell University in 1981. He is a Fellow of the American Mathematical Society (AMS), the Society for Applied and Industrial Mathematics (SIAM) and the Institute for Operations Research and the Management Sciences (INFORMS).

%\vspace{1ex}
%\noindent Biographies and photographs of the other authors are not available.

\listoffigures
\listoftables

\end{spacing}
\end{document}